\documentstyle[amssymb,aps,prl,floats,graphicx]{revtex}
\begin{document}

\twocolumn[\hsize\textwidth\columnwidth\hsize\csname
@twocolumnfalse\endcsname

\title{Vortex lines or sheets -- what is formed in dynamic drives? }

\author{
  V.~B.~Eltsov$^{*,\dagger}$, R.~Blaauwgeers$^{*,\ddagger}$,
  N.~B.~Kopnin$^{*,\S}$,
  M.~Krusius$^*$, J.~J.~Ruohio$^*$,
  R.~Schanen$^{*,\|}$,\\ and E.~V.~Thuneberg$^{*,\S\S}$
}

\address{
  $^*$Low Temperature Laboratory, Helsinki University of Technology,
  P.O.Box 2200, FIN-02015 HUT, Finland\\
  $^\dagger$Kapitza Institute for Physical Problems, Kosygina 2, 117334
  Moscow,  Russia\\
  $^\ddagger$Kamerlingh Onnes Laboratory, Leiden University, P.O.Box
9504,
  2300 RA Leiden, The
  Netherlands\\
  $^\S$Landau Institute for Theoretical Physics, Kosygina 2, 117334
  Moscow, Russia\\
  $^\|$Department of Physics, Royal Holloway University of London,
  Egham, Surrey, TW20 0EX, UK\\
$^{\S\S}$ Department of Physical Sciences, P.O. Box 3000, 90014
University of Oulu, Finland }
\date{\today}
\maketitle
\begin{abstract}
In isotropic macroscopic quantum systems vortex lines can
be formed while in anisotropic
systems also vortex sheets are possible. Based on measurements of
superfluid $^3$He-A, we present the principles
which select between these two competing forms of quantized
vorticity: sheets displace lines if the frequency of the external
field exceeds a critical limit.
The resulting topologically stable state consists of
multiple vortex sheets and has much faster dynamics than
the state with vortex lines.
\end{abstract}
\pacs{05.07.Fh, 74.60.-w, 67.57.Fg, 98.80.Cq}

]

\narrowtext

The state of superfluid $^4$He in rotation was originally
explained by Onsager \cite{Onsager} and Landau and Lifshitz
\cite{LandauLifshitz} by concentrating the required velocity
circulation into co-axial sheets which, when viewed on large
scales, give rise to solid-body-like rotation. Later, Feynman
\cite{Feynman} showed that sheets are not stable in an isotropic
superfluid and that quantized vortex lines are formed instead. The
same explanation was provided by Abrikosov \cite{Abrikosov} for
type II superconductors in a magnetic field.

Recently, however, the vortex sheet has staged a come back: It was
theoretically suggested for unconventional superconductors
\cite{Sigrist} and experimentally verified in the anisotropic
p-wave superfluid $^3$He-A \cite{Parts}. The sheets are formed
from a topologically stable domain-wall-like planar defect into
which the vorticity is confined. Since these quantum systems also
support vortex lines, the controversy remains: What determines the
formation of lines or sheets, or alternatively, what is the
difference in the macroscopic physical properties of the system
when there are lines or sheets? This question is not limited to
superconductors or $^3$He-A, but is of importance for defect
formation in general, possibly in the early universe, in neutron
stars, Bose-Einstein condensates, etc. -- wherever non-trivial
symmetry breaking into a multi-component order-parameter field may
occur.

\begin{figure}[t]
\centerline{\includegraphics[width=0.8\linewidth]{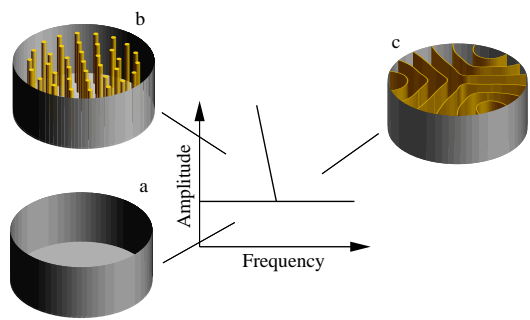}}
\medskip
\caption[RotStates]{The different topologically distinct states of
rotating $^3$He-A depend on the amplitude and frequency of the
drive: (a) The vortex-free Meissner state exists below a critical
value of the amplitude. (b) The state with quantized vortex lines
is created when the drive increases sufficiently slowly. (c)
Vortex sheets are created at higher frequencies. The critical
values depend on the global order-parameter texture and display
large metastability. } \label{RotStates}
\end{figure}

The answer whether lines or sheets are formed depends, according
to our measurements of $^3$He-A, on the applied field: If its
frequency and amplitude exceed the appropriate critical limits,
sheets appear and displace lines, as shown schematically in
Fig.~\ref{RotStates}. The measured phase diagram is supported by
our analysis of the dynamic properties of vortex sheets, and is
further confirmed by numerical simulations. The transition from
lines to sheets is driven by the much faster dynamics of states
with multiple vortex sheets. The roots of this phenomenon share
common features with self organization and pattern formation in
classical systems.

%%%%%%%%%%%%%%%%%%%%%%%%%%%%%%%%%%%%%%%%%%%%%%%%%%%%%%%%%%%%%%
%%%%%%%%%%%%%%%%%%%%%%%%%%%%%%%%%%%%%%%%%%%%%%%%%%%%%%%%%%%%%%%

{\it Vortex lines vs sheets.}-- Superconductors respond to a
magnetic field by producing screening currents. Similarly, a
superfluid attempts to corotate at the container's angular
velocity $\mathbf \Omega$. The two motions resemble each other and
usually take the form of quantized vortex lines so that the
superfluid vorticity $\nabla\times {\mathbf v}_s$ is non-zero only
within vortex cores, having on average the solid-body value
$2{\mathbf \Omega}$. In a state with sheets the irrotational
superflow is channelled parallel to a sheet, while ${\bf v}_s$
jumps across the sheet leaving the non-zero vorticity within the
sheet. The distance between the sheets is $b=\left(3\sigma /\rho
_s\Omega^2\right)^{1/3}$, where $\sigma$ is the surface tension of
the sheet and $\rho_s$ the superfluid density
\cite{LandauLifshitz}. In equilibrium, vortex lines in $^3$He-A
have a lower energy than vortex sheets \cite{Karimaki}.

The stability of the vortex sheet rests on the existence of a
two-fold degeneracy of the ground state, so that a topologically
stable domain wall can exist. In $^3$He-A, it separates
energetically equivalent states with parallel and antiparallel
orientations of the orbital $\hat{\bf l}$ and spin $\hat{\bf d}$
anisotropy directions. In superconductors with broken
time-reversal symmetry the degeneracy could arise from the locking
of $\hat{\bf l}$ to the crystal-field anisotropy \cite{Sigrist}.
The vorticity in the domain wall is quantized \cite{Parts}; it is
distributed along the wall with a period $p=n \kappa /(2\Omega
b)$. Here $\kappa$ is the circulation quantum and $n$ the
topological charge which depends on the structure of the
vorticity. For the vortex sheet in $^3$He-A $n=1$ while separate
vortex lines are generally doubly quantized ($n=2$)
\cite{Blaauwgeers}.

The large-scale structure of the vortex sheet may take
topologically distinct configurations with different numbers $M$
of separated sheets. The minimum is one continuous sheet ($M=1$)
in the form of a double spiral centered around the container axis
(see bottom-right inset in Fig.~\protect\ref{Loops}); this is the
lowest energy state of a vortex sheet at constant $\Omega$. This
configuration is closest to the concentric vortex sheets
originally envisioned by Onsager \cite{Onsager} and by Landau and
Lifshitz \cite{LandauLifshitz}. The other extreme is the maximum
number of sheets, $M_{\rm max}=\pi R/b$, which corresponds to one
connection with the lateral sample boundary per each inter-layer
distance $b$ along the perimeter of the sample ($M_{\rm max}
\approx 20 $ at $\Omega=1\,$rad/s). In the absence of pinning,
like in $^3$He-A, the sheets fold to fill the sample evenly since
a uniform distribution of vorticity is energetically preferred.
Mutual repulsion between the sheets and the boundary conditions
force the sheets to be oriented perpendicular to the boundary.
This promotes their radial orientation, as depicted in
Fig.~\ref{RotStates} (c).

{\em Dynamics of lines and sheets.}-- A time-dependent rotation
drive $\Omega (t)$ exerts a force on the circulation quanta which
is mostly radial in solid-body-like rotation. The vorticity
averaged over the sample, $\Omega_v=\langle|\nabla \times{\bf
v}_s|\rangle/2$, obeys \cite{Sonin}
\begin{equation}
d\Omega _v/dt +2\alpha \Omega _v\left(\Omega _v -\Omega\right)=0,
\label{dOmega/dt}
\end{equation}
where $\alpha$ is the mobility in the radial direction determined
by the friction from the normal excitations. Owing to the strongly
anisotropic order-parameter orientations in the sheet, the
mobility of vorticity in the sheet is also anisotropic. It is
determined by the spatial inhomogeneity of the orbital anisotropy
vector $\hat {\bf l}$ \cite{MutualFriction}:
\begin{equation}
\alpha _i^{-1} =\alpha _0 ^{-1}\int d^2r \left( \nabla _i\hat{\bf
l}\right)^2\;\; .
\end{equation}
Here $i$ refers to the directions along $\parallel$ or
perpendicular $\perp$ to the sheet, and $\alpha _0=\rho _n \rho _s
\kappa /(2\rho \mu)$ is related to the Cross--Anderson orbital
viscosity $\mu$ \cite{VW}. We neglect an order-of-one anisotropy
of $\rho _s$. One obtains $\alpha _\parallel \sim \alpha _0 (p/s)$
and $\alpha _\perp \sim \alpha _0(s/p)$, where $s\approx 40\,\mu$m
is the thickness of the sheet. The mobility of separate vortex
lines is $\alpha _v\sim \alpha _0$, since the dimensions of the
vortex core are $\sim s$ in all directions. Thus the mobility
along the sheet $\alpha _\parallel$ is the highest: $\alpha
_\parallel > \alpha _v > \alpha _\perp$. At $\Omega = 1$ rad/s one
has $\alpha_{\parallel} : \alpha_{v} :\alpha_{\perp} \sim 4 :1 :
1/4$. Since it is the radial mobility that matters in
Eq.~(\ref{dOmega/dt}), radial orientation of the sheets improves
their dynamic response.

For the dynamics also the creation and annihilation of vorticity
is crucial. The circulation quanta are created at a critical
velocity which depends on the order-parameter texture
\cite{Ruutu}. For vortex sheets, vorticity is created at the
connection lines of the sheets with the boundary where the
order-parameter field is already distorted and the critical
velocity is reduced compared to vortex lines \cite{Parts}. Thus
multiple sheets with their $2M$ connection lines provide a large
number of effective nucleation sites. These inject new vorticity
much more efficiently than the competing alternative, one
nucleation center of vortex lines close to the lateral sample
boundary \cite{Blaauwgeers}.

To conclude, the different rates of dynamic response, which
control relaxation towards the instantaneous equilibrium, are
expected to govern the transition between lines and sheets: A
high-frequency drive at sufficient amplitude should replace all
other forms of vorticity with radially oriented multiple vortex
sheets. The transition can occur in the following way: At high
frequency the order-parameter texture around a nucleation center,
where vortex lines are periodically nucleated and annihilated, may
become locally unstable and the beginning of a domain-wall defect
may be created. When existing vortex lines have been annihilated
during a subsequent decrease in the drive, the new vorticity may
now re-enter during the next increase in the drive via the
domain-wall defect, where the critical velocity is reduced and the
mobility is enhanced. A single sheet is unstable in an oscillating
drive and, as discussed later, this leads to the formation of a
state with multiple sheets. A configuration with a given number of
sheets is topologically stable and will persist as a metastable
state if the time-dependent component in the drive is switched off
and $\Omega$ is kept constant.

{\it  Measurement of topological transitions.}-- Our measurements
on $^3$He-A have been performed on a cylindrical sample with a
radius $R = 2\,$mm in an axial magnetic field of 9.6 mT, pressure
33\,bar, and temperature 1.7\,mK, by rotating
the entire refrigerator. NMR spectroscopy is
used to identify the various order-parameter
topologies. Vortex lines or sheets produce characteristic NMR
satellite peaks, while in the vortex-free state there is
no satellite. Although the satellites may overlap (Fig.~2,
insert), they have different dependencies on $\Omega$ which
distinguishes between lines and sheets
\cite{Parts}. The rotation drive is thus used both for generating
transitions between the states and as a
detection tool.

%%%%%%%%%%%%%%%%%%%%%%%%%%%%%%%%%%%%%%%%%%%%%%%%%%%%%%%%%%%%%%%%
\begin{figure}[t]
  \includegraphics[width=0.9\linewidth]{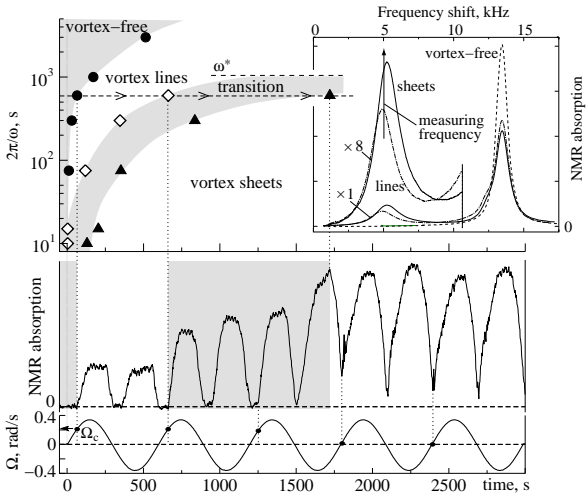}
\medskip
\caption{ {\em (Top)} Sequence of transitions (vortex-free
$\rightarrow$ vortex lines $\rightarrow$ vortex sheets), when the
drive $\Omega(t) = \Omega_1 \sin \omega t$ (with $\Omega_1 =
0.35\,$\,rad/s) is switched on: ($\bullet$) first appearance of
lines, ($\lozenge$) transition to sheets starts,
($\blacktriangle$) transition to sheets is completed. The vertical
axis is the period $2\pi/\omega$ of the drive while the horizontal
axis is the common time axis $(t)$ with the bottom panel. {\em
(Bottom)} Height of the vorticity satellites as a function of time
after switching on the drive $\Omega(t)$, with $2\pi/\omega =
600\,$s which corresponds to the path along the dashed arrow in
the top panel. The signal is recorded at the NMR frequency shown
in the insert. The drive $\Omega(t)$ is plotted below the
satellite signal. {\em (Insert)} NMR absorption spectra of the the
rotating states with lines and sheets at $\Omega = 2.4$\,rad/s and
of the vortex-free state ($\Omega < \Omega_c$).}
  \label{VorSheetTrans}
\end{figure}
%%%%%%%%%%%%%%%%%%%%%%%%%%%%%%%%%%%%%%%%%%%%%%%%%%%%%%%%%%%%%%%%%

The lower part of Fig.~\ref{VorSheetTrans} illustrates transitions
between the three rotating states of Fig.~\ref{RotStates} as a
function of time $t$ when the drive $\Omega(t) = \Omega_1 \sin
\omega t$ is switched on. The starting state is vortex free with
no satellite. Vortex lines begin to form when $\Omega$ reaches the
critical value $\Omega_{c}\approx 0.24$\,rad/s during the first
half period. Here the height of the satellite is proportional to
the {\it total number} of vortex lines. The characteristic
features are two plateaus during one half cycle: the first at zero
height when $|\Omega|$ increases from 0 to $\Omega_c$, and the
second at maximum absorption when $|\Omega|$ decreases from
$\Omega_1$ to roughly $\Omega_1-\Omega_c$. The plateaus correspond
to zero and to maximum number of vortex lines, respectively.

After one full period the conversion to vortex sheets starts. The
response gradually changes and acquires features characteristic to
the vortex sheet: The satellite height grows larger and the
plateaus disappear. For a sheet, the signal height is proportional
to its {\it total length.} The critical velocity is now
vanishingly small which ensures the smooth growth and shrinkage of
the sheets as a function of $\Omega$. Also, the satellite height
does not reach zero when $\Omega$ changes sign because the
backbone of the sheet -- the domain wall -- does not have time to
drift away and continues to contribute to the signal even in the
absence of circulation at $\Omega =0$. These features can be
studied in detail, by switching off the modulation of the drive at
any moment. Thereafter $\Omega$ can be kept at constant value, and
the entire NMR absorption spectrum can be recorded to determine
the frequency shift and height of the satellite.

The upper left panel of Fig.~\ref{VorSheetTrans} illustrates how
long it takes to complete the transitions at different $\omega$.
Note that there exists a characteristic frequency $\omega^*$ such
that $2\pi/\omega^* \sim 10^3\,$s which separates lines and sheets
in the limit $t \rightarrow \infty$: At $\omega < \omega^*$ a
response in terms of lines is stable while at $\omega
> \omega^*$ sheets appear. The conditions in the lower panel
 correspond to the case when $\omega
>\omega ^*$ and sheets are stable. The larger the drive frequency
$\omega$, the faster the transition to sheets starts. Finally,
with $2\pi/\omega \lesssim 20\,$s the transition to sheets starts
directly from the vortex-free state, with no indication of a
signal from vortex lines in between. With increasing drive
amplitude $\Omega_1$ the transition to sheets $\omega^*$ moves to
lower frequencies and is completed faster. Thus the transition
line $\omega ^*(\Omega _1)$ is tilted as shown in
Fig.~\ref{RotStates}.

{\it Identification of sheet topologies.}-- Different vortex-sheet
configurations can be studied by measuring the $\Omega$-dependence
of the NMR satellite. This can be recorded either in the
adiabatically slow limit (Fig.~\ref{Loops}) or as a dynamic response
(Fig.\ \ref{Steps}). To analyze the response of a state with
multiple sheets it is instructive to compare it to that of the
well-known configuration with a single continuously coiled sheet.
The latter provides a convenient reference which can be grown
adiabatically from a single seed of domain wall, by slowly
increasing $\Omega$ from zero \cite{Parts}.

%%%%%%%%%%%%%%%%%%%%%%%%%%%%%%%%%%%%%%%%%%%%%%%%%%%%%%%%%%%%%%%%
\begin{figure}[t]
\centerline{\includegraphics[width=0.8\linewidth]{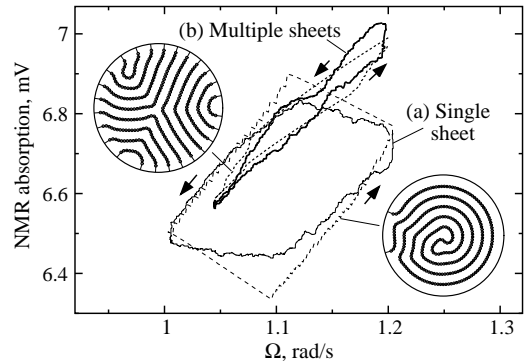}}
\medskip
\caption[Loops]{Measured hysteresis loops in satellite-peak height
when $\Omega$ is slowly changed (solid lines). (a) An
adiabatically grown single sheet. (b) Multiple sheets created from
the single sheet by subjecting it to high-frequency modulation
[$\Omega (t) = (0.4\, \sin \omega t + 1.2)\,$rad/s, with
$2\pi/\omega = 10\,$s]. The broken lines represent numerical
simulations for the configurations shown in the respective
inserts. The only fitting parameter is the coefficient of
proportionality which connects the total length of the sheets to
NMR absorption.} \label{Loops}
\end{figure}
%%%%%%%%%%%%%%%%%%%%%%%%%%%%%%%%%%%%%%%%%%%%

Let us calculate the total sheet length $L$ for a given number of
sheets $M$ and circulation quanta $N$. The vorticity is isolated
from the lateral sample boundary by an annular vortex-free
superflow and confined within a radius $R_v = \sqrt{\kappa N/2\pi
\Omega}$. These central sections of the sheets amount to a total
length of $\pi R_v^2 / b$. At the sample boundary the $2M$ ends of
the $M$ vortex sheets cross the vortex-free region of width $d = R
- R_v \ll R$. We have
\begin{equation}
  L = \pi R_v^2 / b + 2 M d
  \approx \pi R^2/b + 2 d\, ( M - M_{\rm max})\, .\label{length}
\end{equation}

In Fig.~\ref{Loops}, loop (a) is traced by a single sheet ($M=1$):
(i) In the lower branch of the loop which starts from 1.1\,rad/s
in increasing $\Omega$, the counterflow velocity at the boundary
is at the critical value and new vorticity is created while $d$ is
almost constant. We find from Eq.~(\ref{length}) that $L\approx
\pi R_v^2/b\propto N\Omega ^{-1/3}$, if we neglect the small term
$2Md$. Since $N\propto \Omega$, the length is $L\propto
\Omega^{2/3}$ and $dL/d\Omega \approx (2/3)L/\Omega >0$, which
agrees with the positive slope above 1.1\,rad/s. (ii) During
decreasing $\Omega$, $N$ first remains constant, before the
annihilation threshold is reached. In this branch $L\propto \Omega
^{-1/3}$ so that $dL/d\Omega \approx -(1/3)L/\Omega <0$. This
again agrees with the almost twice smaller negative slope when
$\Omega$ first decreases below 1.2\,rad/s. The out-of-phase
response ($L$ increases while the drive $\Omega$ decreases) is the
most distinct characteristic of the single-sheet response at
constant number of circulation quanta $(\Omega_1 \leq \Omega_c$)
\cite{vslt}. (iii) At the annihilation threshold (branch with
$\Omega \lesssim 1.1\,$rad/s, $d\Omega/dt < 0$), vorticity is
expelled, $d\approx {\rm const}$, and again $dL/d\Omega \approx
(2/3)L/\Omega >0$. (iv) Finally, the negative slope of
$dL/d\Omega$ is resumed in the last leg of the loop where $\Omega$
is increased back towards the critical velocity.

In contrast multiple sheets, Fig.~\ref{Loops} loop (b), with
$M\approx M_{\rm max}$, have always $L\propto b^{-1}$, as seen
from Eq.~(\ref{length}), and $dL/d\Omega \approx (2/3)L/\Omega >0$
in agreement with the measured narrow hysteresis loop. The
in-phase response dominates the output from states with multiple
sheets. The prominent difference in the responses of the two sheet
configurations in Fig.~\ref{Loops} has thus a simple explanation
in terms of one adjustable variable $M$.

%%%%%%%%%%%%%%%%%%%%%%%%%%%%%%%%%%%%%%%%%%%%%%%%%
\begin{figure}[t]
\centerline{\includegraphics[width=0.8\linewidth]{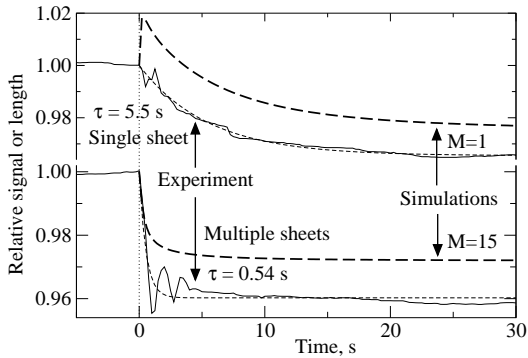}}
\medskip
\caption[Steps]{Dynamic responses of  two different vortex sheet
  configurations  to a step
  change in rotation velocity $\Delta \Omega =-0.15$ rad/s (at $t=0$), with
  $\Omega =1.25$ rad/s initially.  The satellite-peak heights (solid lines)
  normalized to the initial values are recorded as functions of time $t$.
  The thin broken lines show exponential fits to determine the response
  time $\tau$.
  The thick dashed lines show the results from numerical simulations
  ($\sigma/\rho_s = 1.1\cdot 10^{-8}\,$cm$^3\,$s$^{-2}$,
  $\alpha_0 = 30$). The initial increase in the uppermost curve is
  the out-of-phase response of the single sheet (at $N = {\mathrm
  const}) $, which becomes
  experimentally visible only at higher drives $\Omega \gtrsim
1.8\,$rad/s \cite{vslt}.} \label{Steps}
\end{figure}
%%%%%%%%%%%%%%%%%%%%%%%%%%%%%%%%%%%%%%%

{\it Dynamic response.}-- To illustrate the faster dynamics of a
state with multiple sheets, Fig.\ \ref{Steps} provides a
comparison to the single sheet. The responses are recorded to a
step-like change in the rotation drive. The vorticity relaxes with
time constant $\tau=(2\alpha \Omega)^{-1}$ which is an order of
magnitude smaller for the multiple-sheet state than for the single
sheet.

{\it Simulation.}-- These and other measurements can be compared
to numerical simulation, by extending the model in
Ref.~\cite{Heinila} to dynamical processes with an anisotropic
mobility \cite{tobepublished}.  In comparing the responses of
different configurations to slow changes in $\Omega$
(Fig.~\ref{Loops}), it appears that the most important factor is
the number of sheets, whereas the detailed arrangement of the
sheets is less influential. Our dynamical simulations confirm that
in an oscillating drive, a single sheet transformes into multiple
sheets: When $\Omega$ increases rapidly azimuthal sections of the
sheet become unstable. When the number of circulation quanta
grows, an azimuthal outer sheet becomes overfilled as the quanta
have no time to move inside the sheet. This section then becomes
corrugated and a bulge is formed, to redistribute the vorticity
more evenly. During the subsequent decrease of $\Omega$ the
vorticity expands and the bulge may touch the sample boundary.
This produces two new connections and improves the radial
alignment of the sheets. Ultimately, the faster motion of
vorticity in such sheets suppresses the formation of new bulges
when $M$ has grown to $(0.6 \div 0.8)M_{\rm max}$. In
Fig.~\ref{Steps} the step response of such a dynamically created
sheet configuration is compared to that of the single sheet from
which it was created. In both cases the measured curves are in
reasonable agreement with the calculations.

{\it Conclusion.}-- Superfluid $^3$He-A provides the first example
of a quantum system where the temporal properties of the external
field determine in what form of quantized vorticity the response
occurs. As a function of frequency, a topological transition takes
place from linear to planar order-parameter configurations. Among
all possible states of vorticity in $^3$He-A, a state with
radially oriented multiple vortex sheets is established that
provides the fastest dynamic response at high frequencies. Many of
the features, which make this phenomenon possible are also present
in other systems with a multi-component order parameter, and thus
this could be a generic property of quantum systems with intrinsic
anisotropy.

We thank E. Sonin and G. Volovik for instructive discussions.
This work was funded by the EU-IHP program and by the
Russian Foundation for Basic Research.

\end{document}